\documentclass[9pt]{article}
\usepackage{spconf,amsmath,graphicx}
\usepackage{multirow}
\usepackage{multicol}
\usepackage{tabularx}
\usepackage{floatrow}
\usepackage{url}
\usepackage{xcolor}

\title{ICASSP 2021 JOINT GENDER AND AGE ESTIMATION BASED ON SPEECH SIGNALS USING X-VECTORS AND TRANSFER LEARNING}
\name{Author(s) Name(s)}
  \name{D. Kwasny, D. Hemmerling}
	\address{AGH University of Science and Technology, 
	Department of Measurement and
Electronics\\
	al. Mickiewicza 30 30-059 Kraków}
\begin{document}
%
\maketitle
\begin{abstract}
In this paper we extend the x-vector framework for the task of speaker's age estimation and gender classification. 
In particular, we replace the baseline multilayer-TDNN architecture with QuartzNet, a convolutional architecture that has gained success in the field of speech recognition.
We further propose a two-staged transfer learning scheme, utilizing large scale speech datasets: VoxCeleb and Common Voice, and usage of multitask learning to allow for joint age estimation and gender classification with a single system.
We train and evaluate the performance on the TIMIT dataset. The proposed transfer learning scheme yields consecutive performance improvements in terms of both age estimation error and gender classification accuracy and the best performing system achieves new state-of-the-art results on the task of age estimation on the TIMIT TEST dataset with MAE of 5.12 and 5.29 years and RMSE of 7.24 and 8.12 years for male and female speakers respectively while maintaining a gender classification accuracy of 99.6\%.
\end{abstract}

\section{Introduction}
\label{sec:intro}
Speech plays important role in the interpersonal communication. In addition to the content of the speech, it also contains the information about speaker's identity, emotions, gender, origin, etc. The creation of a system that automatically and with the greatest possible precision could estimate the speakers' data such as gender and age is desired by companies using telephony to communicate with customers. 

\textbf{Related work:} In recent years the DNN usage for speech processing became very efficient in the identification of relevant information for classification and prediction of gender and age. 
The paper \cite{a9} presents the usage of deep bottleneck extractor and a GMM–UBM classifier for speaker age and gender classification. The overall accuracy achieved by the proposed classification system is 57,63\%. Authors of \cite{spain} proposed a convolutional-recurrent neural network architecture for gender and age prediction from speech. Gender is recognized with an average error below 1,55\%, while the probability of speakers' age to be correctly classified into three age groups is higher than 80\% on average. The paper of \cite{hebbar1} presents the implementation of DNN with a transfer learning strategy in a convolutional layer based network for gender classification from movie audio data. The authors achieved 85\% weighted accuracy in the best set up. The authors of~\cite{ghahremani2018end} proposed usage of the x-vector neural network architecture, which has been shown to offer state-of-the-art results in the realm of speaker recognition~\cite{villalba2019state}, to the age estimation task. The reported mean absolute error (MAE) was 4,92 years on the NIST SRE10 dataset. In a recent paper from 2019 \cite{icassp2019}, the authors proposed a novel, unified DNN architecture for a joint height/age estimation system that let them improve over the baseline support vector regression solution by at least 0.6 years in terms of the root mean square error (RMSE). In the case of age estimation, the RMSE errors are 7.60 and 8.63 years for males and females respectively, evaluated on the TIMIT database. The same database was used in the research conducted by authors of \cite{paper2020}. They explored different features that stream for age and body build estimation derived from the speech spectrum using support vector regression. The MAE of age estimation in this approach is 5.2 and 5.6 years for males and females respectively.

\textbf{Contribution:} In this work we propose an implementation of x-vector-based DNN system for joint age estimation and gender classification. 
In particular, the proposed system uses different embedding network architecture (QuartzNet~\cite{kriman2020quartznet}), compared to the previous attempts presented in~\cite{ghahremani2018end} and~\cite{paper2020.2}, which used the TDNN network. 
Moreover, unlike the system from~\cite{ghahremani2018end}, which can only be used to estimate the age of the speaker, and~\cite{paper2020.2}, which performs joint age/gender classification, but only distinguishes between 4 age classes, our system learns to jointly estimate the exact age of the speaker and classify the gender.
Furthermore, we propose to use transfer learning from speaker recognition and age estimation/gender classification on different datasets. 
The proposed transfer learning scheme yields consecutive results improvements and the system achieves new state-of-the-art results on the TIMIT dataset.
\section{METHODS}

The general idea behind the x-vector architecture is a usage of a few convolutional layers responsible for capturing local dependencies, and a pooling layer which computes statistics (mean and standard deviation) over the time dimension. 
The result of the pooling layer is then passed through additional affine layer to form the final embedding vector, which is of fixed size regardless of the length of the input sequence. 
Recent results in the field of speaker recognition~\cite{villalba2019state} shows, that the performance of the x-vector-based systems is reliant on the architecture of the embedder network used and deeper architectures such as TDNN-F~\cite{povey2018semi} and ResNet~\cite{he2016deep} offer performance gains when compared to the baseline TDNN-based system. 

\begin{figure}[H]
    \centering
    \includegraphics[width=0.7\textwidth]{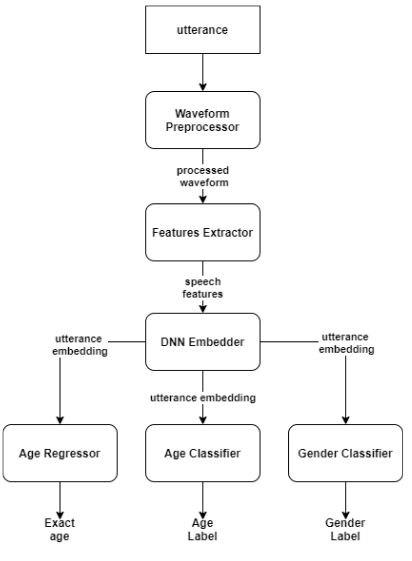}
    \caption{High level representation of the system for joint age estimation and gender classification.}
    \label{fig:scheme}
\end{figure}

Encouraged by these results, we decided to follow with the x-vector framework for the joint gender classification and age estimation task.
The scheme of the proposed system is shown in figure~\ref{fig:scheme}.

We distinguish two groups of modules making-up the overall system: the embedder, which given a variable-length audio sequence outputs a fixed-size embedding of the input utterance and the front-end classifier and regressor networks, which given the embedding perform the classification/regression task. 
Instead of the TDNN-based embedder used in~\cite{ghahremani2018end}, we used a deeper network, which is heavily based-off the QuartzNet~\cite{kriman2020quartznet} architecture.

\begin{table}[H]
\begin{tabular}{|c|c|c|c|c|}
\hline
\textbf{\begin{tabular}[c]{@{}c@{}}Block \\ Name\end{tabular}}                          & \textbf{\begin{tabular}[c]{@{}c@{}}Kernel \\ Length\end{tabular}} & \textbf{Repeats} & \textbf{Residual} & \textbf{Size}                                        \\ \hline
\textbf{Input}                                                                          & 3                                                                 & 1                & TRUE              & 512                                                  \\ \hline
\textbf{Block\_1}                                                                       & 5                                                                 & 2                & TRUE              & 512                                                  \\ \hline
\textbf{Block\_2}                                                                       & 7                                                                 & 2                & TRUE              & 512                                                  \\ \hline
\textbf{Block\_3}                                                                       & 9                                                                 & 2                & TRUE              & 512                                                  \\ \hline
\textbf{Final}                                                                          & 1                                                                 & 1                & FALSE             & 1500                                                 \\ \hline
\textbf{\begin{tabular}[c]{@{}c@{}}Stats \\ pooling\\  (Mean  \\ Std)\end{tabular}} & -                                                                 & 1                & FALSE             & \begin{tabular}[c]{@{}c@{}}1500+\\ 1500\end{tabular} \\ \hline
\textbf{\begin{tabular}[c]{@{}c@{}}Dense\\  BatchNorm \\ ReLu\end{tabular}}    & -                                                                 & 1                & FALSE             & 512                                                  \\ \hline
\textbf{\begin{tabular}[c]{@{}c@{}}Dense \\  BatchNorm \\  ReLu\end{tabular}}    & -                                                                 & 1                & FALSE             & 512                                                  \\ \hline
\end{tabular}
\caption{QuartzNet embedder architecture.}
\label{tab:quartznetXvec}
\end{table}

The QuartzNet architecture we used is composed of a convolutional layer followed by a sequence of 3 groups of blocks. 
The blocks in a group are identical and are repeated 2 times.
There is one additional convolutional layer proceeded with a pooling layer, which aggregates the mean and standard deviation across the time dimension.
The output of the pooling layer is then passed through a stack of two affine layers to form the final embedding.
The embedder network architecture is summarized in table~\ref{tab:quartznetXvec}.
The details of the front-end modules are presented in table~\ref{tab:xvectorFrontend}.
\begin{table}[H]
\begin{tabular}{|c|c|c|c|}
\hline
\textbf{Network}                                                                       & \textbf{Layer}                                                         & \textbf{Input SIze} & \textbf{Output SIze} \\ \hline
\multirow{2}{*}{\textbf{\begin{tabular}[c]{@{}c@{}}Binary \\ Classifier\end{tabular}}} & \begin{tabular}[c]{@{}c@{}}Dense + \\ ReLu + \\ BatchNorm\end{tabular} & 512                 & 512                  \\ \cline{2-4} 
                                                                                       & \begin{tabular}[c]{@{}c@{}}Dense + \\ Sigmoid\end{tabular}             & 512                 & 1                    \\ \hline
\multirow{2}{*}{\textbf{\begin{tabular}[c]{@{}c@{}}Multi \\ Classifier\end{tabular}}}  & \begin{tabular}[c]{@{}c@{}}Dense + \\ ReLu + \\ BatchNorm\end{tabular} & 512                 & 512                  \\ \cline{2-4} 
                                                                                       & \begin{tabular}[c]{@{}c@{}}Dense + \\ Softmax\end{tabular}             & 512                 & 8                    \\ \hline
\multirow{2}{*}{\textbf{Regressor}}                                                    & \begin{tabular}[c]{@{}c@{}}Dense + \\ ReLu + \\ BatchNorm\end{tabular} & 512                 & 512                  \\ \cline{2-4} 
                                                                                       & Dense                                                                  & 512                 & 1                    \\ \hline
\end{tabular}
\caption{Front-end modules details.}
\label{tab:xvectorFrontend}
\end{table}

One of the most challenging aspects of using more complex DNN architectures is the difficult training process, especially when the amount of training data is limited and the network tends to overfit the training data. 
To fight these limitations we have decided to use two techniques which has found great success in various areas of deep learning: transfer learning and multitask learning.
In particular, we explored various pretraining schemes involving pretraining the embedder for the task of speaker recognition or age/gender classification but on a different dataset with more data as well as a combination of both. 
As for the multitask learning, the proposed system jointly learns to predict both the gender and the age of the speaker. 
This is beneficial in two ways, as only a single system needs to be trained to predict multiple speaker characteristics and the front-end networks may benefit from the implicit information about the gender and age of the speaker conatined in the embeddings.
We also employ an additional cross-entropy loss for the classification objective for age-group classification, as it has been shown to stabilize the training of the age estimator~\cite{ghahremani2018end}.

\subsection{Data}
For the purpose of this research we have used 3 datasets: VoxCeleb1~\cite{voxceleb}, the Common Voice dataset~\cite{ardila2019common,kaggle_common} and DARPA-TIMIT dataset~\cite{timit}. The VoxCeleb1 dataset was created primarily with the aim of accelerating research in the field of speaker identification and verification. 
It is gender-balanced with around 55\% of male speakers.
We have used this dataset to pretrain the embedder network for the speaker recognition task. 
For the sake of reproducibility we have decided to use the subset of the english part of the CommonVoice dataset available through the Kaggle website~\cite{kaggle_common}. 
We used only those entries, which contained information about both age group and gender of the speaker, further restricting gender to be either male or female to allign with the gender present in the TIMIT dataset. 
In total, it corresponds to approximately 80 hours of data in the train set, and 1.5 hours of data in both validation and test sets. For final training and validation we used the DARPA-TIMIT dataset. A random TRAIN-TEST split is performed on the default TRAIN subset of the data, which corresponds to roughly 3.5 hours of train data and 0.5 an hour of validation data. 
For the final evaluation of results the default TEST set is used, which corresponds to roughly 1500 utterances and 1.5 hour of data, all spoken by speakers not present in the training set.
The usage of the default TEST set let us fairly compare with results obtained by previous works on this dataset.

\subsection{Experimental setup}
Before the feature extraction, the following processing steps were applied to the raw waveforms.
First, Voice Activity Detection was used to remove the non-speech frames from the input utterance and then a random, 5-seconds long crop was extracted, as it has been shown to improve the results of x-vector based system presented in~\cite{ghahremani2018end}.
The cropped signal was then volume-normalized to the common value of decibel relative to full scale (dBFS) of -30 dB~\cite{wan2018generalized}.

Two feature sets commonly used in various speech classification task were considered: MFCCs and mel spectrograms.
In particular we tried 30-dimensional MFCCs features and 64-dimensional mel spectrograms.
In both cases we use 25 ms-long hamming window with a slide of 10 ms, lower cut-off frequency of 40 Hz and upper cut-off frequency of 8000 Hz. 

We use a combination of binary cross entropy loss for gender classification, cross entropy loss for age-group prediction that stabilize the training of the age estimator~\cite{ghahremani2018end} and mean square error loss for the age estimation task
The losses are combined according to weights shown in table~\ref{tab:losses}.
To allow pretraining the regressor on the Common Voice dataset, we approximate the speaker's exact age to be equal to the middle value of the corresponding age bin.

\begin{table}[H]
\begin{tabular}{|c|c|c|}
\hline
\multicolumn{3}{|c|}{Loss weights}                                                                                                                                                                                                      \\ \hline
\begin{tabular}[c]{@{}c@{}}BCE Loss\\ for gender \\classification\end{tabular} & \begin{tabular}[c]{@{}c@{}}CE Loss \\ for age group \\classification\end{tabular} & \begin{tabular}[c]{@{}c@{}}MSE Loss \\ for age \\estimation\end{tabular} \\ \hline
1.0                                                                          & 1.0                                                                             & 0.001                                                                  \\ \hline
\end{tabular}
\caption{Loss weights for joint gender classification and age estimation.}
\label{tab:losses}
\end{table}

\begin{table}[H]
\begin{tabular}{|c|c|} 
\hline
\multicolumn{2}{|c|}{\textbf{Optimizer info}} \\ \hline
\textbf{Name} & NovoGrad~\cite{ginsburg2019stochastic} \\ \hline
\textbf{Learning Rate} & 0,001 \\ \hline
\textbf{Learning Rate Policy} & Cosine Annealing~\cite{cosineann} \\ \hline
\textbf{Weight Decay} & 0,001 \\ \hline
\textbf{Beta 1} & 0,95 \\ \hline
\textbf{Beta 2} & 0,5 \\ \hline
\textbf{Warmup Steps} & 10000 \\ \hline
\textbf{Batch Size} & 16 \\ \hline
\textbf{Num Epochs (CV)} & 100 \\ \hline
\textbf{Num Epochs (TIMIT)} & 50 \\ \hline
\end{tabular}
\caption{Optimizer configuration.}
\label{tab:optimizerConfig}
\end{table}

The network was trained with the optimizer configuration shown in table~\ref{tab:optimizerConfig} on a single nVidia Quadro RTX5000 graphics card.
Before the experiments, all the data was converted to a common \textit{wav} format and down-sampled to the frequency of 16 kHz.
To conduct all the experiments we used the nVidia NeMo framework~\cite{nemo2019}.

\section{RESULTS}
\subsection{Gender classification results}
Results of the gender classification task evaluated on the TIMIT TEST dataset with different pretraining schemes and feature sets are shown in table~\ref{tab:genderResults}.
All networks were pretrained on the datasets shown in the "Pretrained on" column and then trained on the TIMIT TRAIN dataset.
The proposed transfer learning scheme resulted in improved accuracy in all cases, especially the system that uses mel spectrograms benefited from pretraining on the VoxCeleb dataset (it's accuracy raised from 95,40\% to 98,90\%). 
On the other hand, the MFCCs-based system benefited more from the additional pretraining on the Common Voice dataset, achieving the highest gender classification accuracy of 99,60\%.
\begin{table}[]
\begin{tabular}{|c|c|c|c|}
\hline
Features                                                                   & Pretrained on                                                                     & Group  & Accuracy \\ \hline
\multirow{9}{*}{MFCC}                                                      & \multirow{3}{*}{-}                                                                 & all    & 98,30\%  \\ \cline{3-4} 
                                                                           &                                                                                   & female & 97,00\%  \\ \cline{3-4} 
                                                                           &                                                                                   & male   & 98,90\%  \\ \cline{2-4} 
                                                                           & \multirow{3}{*}{VoxCeleb}                                                         & all    & 98,50\%  \\ \cline{3-4} 
                                                                           &                                                                                   & female & 96,40\%  \\ \cline{3-4} 
                                                                           &                                                                                   & male   & 99,50\%  \\ \cline{2-4} 
                                                                           & \multirow{3}{*}{\begin{tabular}[c]{@{}c@{}}VoxCeleb \\ Common Voice\end{tabular}} & all    & \textbf{99,60\%}  \\ \cline{3-4} 
                                                                           &                                                                                   & female & 98,80\%  \\ \cline{3-4} 
                                                                           &                                                                                   & male   & \textbf{100,00\%} \\ \hline
\multirow{9}{*}{\begin{tabular}[c]{@{}c@{}}Mel \\ Spectogram\end{tabular}} & \multirow{3}{*}{-}                                                                & all    & 95,40\%  \\ \cline{3-4} 
                                                                           &                                                                                   & female & 86,40\%  \\ \cline{3-4} 
                                                                           &                                                                                   & male   & 99,90\%  \\ \cline{2-4} 
                                                                           & \multirow{3}{*}{VoxCeleb}                                                         & all    & 98,90\%  \\ \cline{3-4} 
                                                                           &                                                                                   & female & 97,50\%  \\ \cline{3-4} 
                                                                           &                                                                                   & male   & 99,60\%  \\ \cline{2-4} 
                                                                           & \multirow{3}{*}{\begin{tabular}[c]{@{}c@{}}VoxCeleb \\ Common Voice\end{tabular}} & all    & 99,40\%  \\ \cline{3-4} 
                                                                           &                                                                                   & female & \textbf{98,90\%}  \\ \cline{3-4} 
                                                                           &                                                                                   & male   & 99,60\%  \\ \hline
\end{tabular}
\caption{Gender classification results on the TIMIT TEST dataset.}
\label{tab:genderResults}
\end{table}
\subsection{Age classification results}
The results of the age estimation evaluated on the TIMIT TEST dataset are shown in table~\ref{tab:ageResults}.
Similarly as in the gender classification case, the proposed pretraining yields improvements in every scenario.
While the results of the systems with embedder pretrained on the VoxCeleb dataset are very similar, the system that uses MFCCs shows superior performance when no pretraining is applied as well as when additional pretraining on Common Voice is used.
The MFCC-based system pretrained only on VoxCeleb already offers results better then the best results obtained with a DNN-based system on this dataset in terms of the RMSE, presented in~\cite{icassp2019}, while the MAE matches this achieved by the authors in~\cite{paper2020} with a feature-engineering based approach.
Additional pretraining of the whole MFCC-based system on the Common Voice dataset yields another performance improvement and allows us to report the new state-of-the-art on the task of age estimation on the TIMIT TEST dataset with MAE of 5.12 and 5.29 years and RMSE of 7.24 and 8.12 years for male and female speakers respectively.

\begin{table}[H]
\begin{tabular}{|c|c|c|c|c|}
\hline
Features                                                                   & Pretrained on                                                                     & Group  & MAE  & RMSE \\ \hline
\multirow{9}{*}{MFCC}                                                      & \multirow{3}{*}{-}                                                                & all    & 5,98 & 8,47 \\ \cline{3-5} 
                                                                           &                                                                                   & female & 6,28 & 9,42 \\ \cline{3-5} 
                                                                           &                                                                                   & male   & 5,83 & 7,96 \\ \cline{2-5} 
                                                                           & \multirow{3}{*}{VoxCeleb}                                                         & all    & 5,37 & 7,74 \\ \cline{3-5} 
                                                                           &                                                                                   & female & 5,65 & 8,53 \\ \cline{3-5} 
                                                                           &                                                                                   & male   & 5,23 & 7,31 \\ \cline{2-5} 
                                                                           & \multirow{3}{*}{\begin{tabular}[c]{@{}c@{}}VoxCeleb \\ Common Voice\end{tabular}} & all    & \textbf{5,18} & \textbf{7,54} \\ \cline{3-5} 
                                                                           &                                                                                   & female & \textbf{5,29} & 8,12 \\ \cline{3-5} 
                                                                           &                                                                                   & male   & \textbf{5,12} & \textbf{7,24} \\ \hline
\multirow{9}{*}{\begin{tabular}[c]{@{}c@{}}Mel \\ Spectogram\end{tabular}} & \multirow{3}{*}{-}                                                                & all    & 6,4  & 9,12 \\ \cline{3-5} 
                                                                           &                                                                                   & female & 7,38 & 9,9  \\ \cline{3-5} 
                                                                           &                                                                                   & male   & 8,71 & 8,71 \\ \cline{2-5} 
                                                                           & \multirow{3}{*}{VoxCeleb}                                                         & all    & 5,4  & 7,78 \\ \cline{3-5} 
                                                                           &                                                                                   & female & 5,59 & 8,35 \\ \cline{3-5} 
                                                                           &                                                                                   & male   & 5,31 & 7,48 \\ \cline{2-5} 
                                                                           & \multirow{3}{*}{\begin{tabular}[c]{@{}c@{}}VoxCeleb\\ Common Voice\end{tabular}}  & all    & 5,34 & 7,70 \\ \cline{3-5} 
                                                                           &                                                                                   & female & 5,36 & \textbf{7,79} \\ \cline{3-5} 
                                                                           &                                                                                   & male   & 5,32 & 7,65 \\ \hline
\end{tabular}
\caption{Age estimation results on the TIMIT TEST dataset.}
\label{tab:ageResults}
\end{table}
\section{DISCUSSION AND CONCLUSION}
In this work we proposed to replace the basic TDNN-based x-vector architecture with a deeper QuartzNet-based x-vector architecture and to train it for the joint speaker' age estimation and gender classification. 
We proposed a novel, two-stage transfer learning scheme, utilizing data available in large scale speech corpora: VoxCeleb and Common Voice. 
We trained and evaluated the system on the popular TIMIT dataset, using the default TRAIN-TEST split.
The proposed transfer learning scheme resulted in consecutive performance gains in every scenario. 
Using an embedder network that was pretrained on different task (speaker recognition) and dataset (VoxCeleb) we reported a RMSE of 7.31 and 8.53 years and MAE of 5.23 and 5.65 years for male and female speakers respectively.
These results are already better then the best results on this task in terms of RMSE metric (7.60 male/8.63 female) achieved with a DNN-based system~\cite{icassp2019} and on-par in terms of the MAE with the results obtained with a feature engineering-based approach, reported in~\cite{paper2020}.
Additional pretraining on the Common Voice dataset yielded another improvement and the system achieved new state-of-the-art results in terms of both MAE and RMSE metrics, with MAE of 5.12 and 5.29 years and RMSE of 7.24 and 8.12 years for male and female speakers respectively while maintaining a gender classification accuracy of 99.6\% on the TIMIT TEST dataset.
These results further confirm the effectiveness of the proposed approach.



\bibliography{ICASSP_shortened_XARXIV}{}
\bibliographystyle{plain}

\end{document}